\documentclass[amssymb,onecolumn,prb,showpacs]{revtex4}
\usepackage{amsmath} 
\usepackage{amsfonts} 
\usepackage{graphicx} 
\usepackage{epstopdf}
\usepackage{multirow}
\usepackage[a4paper, left=2.5cm, right=2.0cm, top=3.5cm, bottom=3.5cm, headsep=1.2cm]{geometry}
\usepackage{adjustbox} 


\begin{document}
\linespread{1.1} 

\title{Observation of dynamics of hydrogen bonds in {\rm TGS} crystals by means of measurements of pyroelectric currents induced by changes of temperature}
\author{M. Trybus, T. Paszkiewicz and B. Wo{\'s}}

\affiliation{Rzesz{\'o}w University of Technology, ul. Powsta{\'n}c{\'o}w Warszawy 6, 35-959 Rzesz{\'o}w, Poland}
\email[Corresponding author: M. Trybus, email: ]{m_trybus@prz.edu.pl} 
\date{\today}

\begin{abstract}
\small{
The study of pyroelectric response of monocrystalline {\rm TGS} cubic specimens to changes of temperature induced by linear and pulse heating of three mutually perpendicular pairs of cube sides demonstrated a complicated structure of signals. We attribute their forms to the activation of various hydrogen bonds between glycine I, II, III molecules. In the case of pulse heating the pyroelectric signal is observed also in the paraelectric phase. The applied measurement method is relatively simple but precise and can be realized in most of dielectric measurements laboratories.}
\end{abstract}

\maketitle
\section{Introduction}
\label{sect:1}

Triglycine sulfate ({\rm TGS}) is a hybrid organic-inorganic crystal. Due to its ferroelectric properties at room temperature, it is one of the most comprehensively studied ferroelectric materials for infrared, non-cooled thermal detectors. {\rm TGS} is a model uniaxial ferroelectric --- attractive because of its excellent pyroelectric properties and high figures of merits. The material undergoes second-order, order-disorder phase transition at about 49 $^{\circ}$C. Crystals are monoclinic in both polar and non-polar phases. Growth of single crystals is relatively easy, and properties of {\rm TGS} are widely described in numerous papers. 

The primitive vectors of the {\rm TGS} crystalline lattice are denoted by ${\mathbf a}$, ${\mathbf b}$, ${\mathbf c}$. The lattice constants of the {\rm TGS} monoclinic unit cell are $a$=0.15 \AA, $b$=12.69 \AA, $c$=5.73 \AA, and $\beta$=105$^{o}$ (Wood \cite{Wood}). In the ferroelectric phase, the polarization vector is directed along vector $\mathbf{b}$, which is perpendicular to the plane defined by the vectors $\mathbf{a}$ and $\mathbf{c}$ (henceforth we shall refer it to the $b$-side). Vector $\mathbf{b}$ determines the $b$ axis. 

Using various methods, many authors investigated the phase transition and structural changes of {\rm TGS} and its doped crystals. The aim of such investigations is to find structural origin of ferroelectricity and to explain mechanisms of the phase transition in {\rm TGS} family. However, there are still questions about the trigger of phase transition in {\rm TGS}. 

\begin{figure}[ht]
	\centering
	\includegraphics[width=0.5\textwidth, angle=-90]{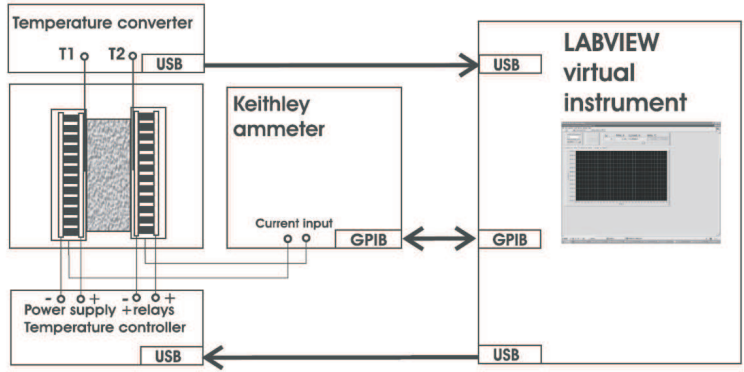}
	\caption{Schematic view of the measuring system.} 
	\label{fig:1}
\end{figure}

Keeping in mind that pyroelectric effect in $b$ axis direction is strong and may veil subtle contributions of hydrogen bonds that may be present in vicinity of the critical temperature T$_c$, we decided to perform pyroelectric measurements in two orthogonal directions perpendicular to the ferroelectric $b$ axis. The way to apply our chosen method of measurements is to use samples of {\rm TGS} in the form of a cube. Such form makes measurements of electric current in three nonplanar directions easy and repeatable under controlled conditions. With the use of the setup described in our papers \cite{Trybus,Wos}, ferroelectric samples can be stimulated by linear changes of temperature and temperature pulses with desired parameters (such as the duration, amplitude, or fill factor of the pulse). We used a pair of micro-Peltier cells for heating or cooling of pairs of parallel sides of cubic or plate shaped samples. The measuring system is shown in Fig. \ref{fig:1}.

The dependence of the pyroelectric current response to changes of temperature of {\rm TGS} was studied by several authors. Chynoweth \cite{Chynoweth} determined the pyroelectric signal as a function of temperature in both ferroelectric and paraelectric phases. He also observed a small pyroelectric signal above the Curie point. Chynoweth used the chopped light beam to generate pyroelectric signals which made the temperature of {\rm TGS} specimens slightly higher than of an oven. White and Wieder induced the paraelectric response in ferroelectric {\rm TGS} using microwaves \cite{White}. Simhony and coworkers used rectangular \cite{Shaulov} and step \cite{Simhony} infrared signals to obtain the temperature dependence of parameters of pyroelectric voltage response, namely of the initial slop, peak value, rise and fall times. Hadni and collaborators scanned the surface of {\rm TGS} plate with a laser beam \cite{Hadni,Lambert}. Measuring pyroelectric current, they observed motion of domain walls as well as location of domain nucleation centers. The domain structure in a plane perpendicular to ferroelectric $b$ axis (henceforth referred to the $b$-side) of {\rm TGS} plates was revealed in experiments of Bhide et al. \cite{Bhide}. These authors induced the pyroelectric signal with a laser beam which can be displaced over the crystal surface. They noted that there are limits to the resolution of small domains related to the size of the laser beam cross-section. Al-Allak and Dewsbery \cite{Al-Allak} studied the pyroelectric coefficient along the ferroelectric axis in a single domain crystal of {\rm TGS}. 

In all of the above described experiments, currents were directed along the $b$-axis. However, Fugiel studied the pyroelectric properties of {\rm TGS} along the non-polar $c$-axis in the {\rm TGS} \cite{Fugiel-a}. In experiments performed by Fugiel et al., {\rm TGS} crystals were placed in a constant electric field perpendicular to the ferroelectric axis \cite{Fugiel-a,Fugiel-b,Cwikiel-a}. 

Pyroelectric signals induced by means of the described methods depend on the topography and on the domain pattern of a surface under study. Hence, high resolution studies of ferroelectric materials are important for better understanding of nucleation and growth of ferroelectric domains. The domain structures of {\rm TGS} were studied by various modes of Atomic Force Microscopy (cf. references given in \cite{Correira,Bluhm,Abplanalp,Orlik}), Electrostatic Force Microscopy \cite{Shin} and Piezoresponse Force Microscopy \cite{Cwikiel-b}. Using results of neutron and X-ray single crystal diffuse scattering, Hudspeth et al. proposed a microscopic mechanism of domains development \cite{Hudspeth,Goossens}. 

We shall underline that our method of measuring pyroelectric currents \cite{Trybus} averages out effects of the surfaces topography as well as the distribution and time dependence of the surface charge. Such effects were observed in experiments described in papers \cite{Correira}$^-$\cite{Cwikiel-b} and in papers cited therein. 

To avoid effects studied in papers \cite{Correira}$^-$\cite{Cwikiel-b} before each measurement our {\rm TGS} samples were heated from about 30 $^{\circ}$C to 55 $^{\circ}$C and then cooled back down to 30 $^{\circ}$C. This way, each initial measurement was performed on aged {\rm TGS} samples while all successive measurements were conducted on freshly rejuvenated samples. We studied the behavior of three cubic samples, but we only present the results for the sample for which the dynamics of the signal measurements was the best. We induced the response of samples to linear changes of their temperature. In the case of heating of $b$-sides we also applied heat pulses. 

\section{Experiments}
\label{sect:2}
Single crystals of {\rm TGS} were grown from an aqueous solution of stoichiometric quantities of the aminoacetic and sulphuric acids, using the static method of water evaporation, as described in numerous publications. Colorless, transparent crystals of good optical quality were obtained. The ferroelectric {\it{b}} axis of the samples was oriented using natural cleavage of single crystals. After mechanical treatment, samples with dimensions of about 5$\times$10$^{-3}$ m (width), 5$\times$10$^{-3}$ m (length) and 5$\times$10$^{-3}$ m (thickness) were fabricated. Silver electrodes were attached to the sample sides. They were used for pyroelectric measurements with the use of a sample holder and measuring system as described in ref \cite{Trybus}.


\subsection{Linear thermal excitation: $b$-sides are heated}
\label{sc:3}
Our former experiments were performed on plate-shaped specimens \cite{Trybus,Wos}. In order to compare the pyroelectric response of cubic and plate-shaped samples, we fabricated the reference sample of dimensions 5$\times$10$^{-3}$ m (width), 
5$\times$10$^{-3}$ m (length) and 2$\times$10$^{-3}$ m (thickness). We used the same single crystal to fabricate both cubic and plate-shaped samples. 

In case of cubic samples, one has to take into account the influence of samples' thickness on the difference of temperatures inside the sample holder and the sample. The volume of cubic samples is about 2.5 times greater than that of plate-shaped ones. In the case of samples of volumes like those used in our experiment, one may expect shifts of measured critical temperature of the phase transition being a result of inertia of heat transfer across samples, especially in case of fast temperature changes. 

\begin{figure}[ht]
	\centering
	\includegraphics[width=0.5\textwidth,angle=-90]{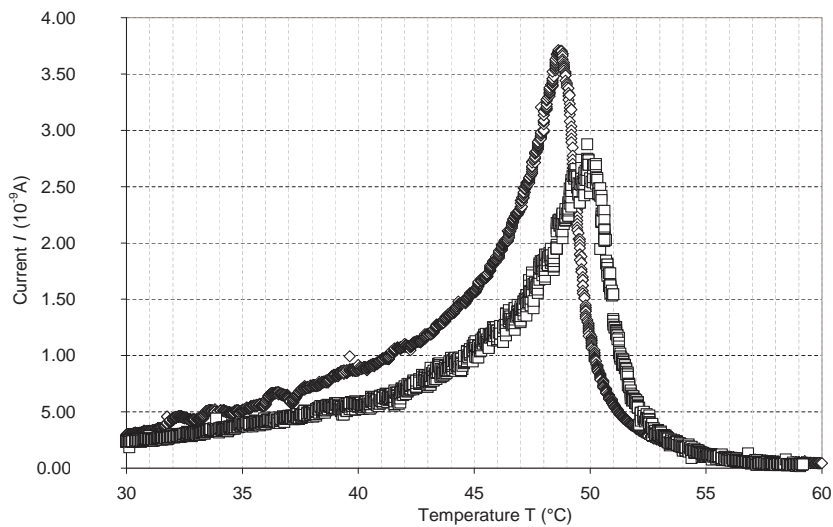}
	\caption{Pyroelectric current of plate-shaped ($\Diamond$) and cubic sample ($\Box$) for linear heating of $b$-sides.} 
	\label{fig:2}
\end{figure}

\begin{figure}[ht]
	\centering
		\includegraphics[width=0.5\textwidth,angle=-90]{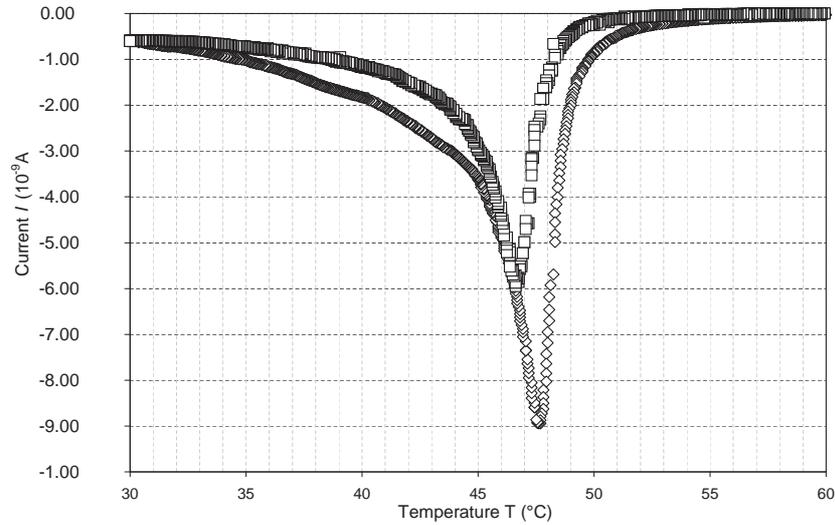}
	\caption{Pyroelectric response of plate-shaped ($\Diamond$) and cubic sample ($\Box$) for linear cooling of $b$-sides.} 
	\label{fig:3}
\end{figure}

Results of measurements obtained for the plate-shaped and cubic samples heated and cooled linearly with rates 1.65 $^{\circ}$C/minute and 5.25 $^{\circ}$C/minute, respectively, are presented in Figs. \ref{fig:2} and \ref{fig:3}.

In the case of linear heating and cooling of our plate-shaped sample, pyroelectric current has a maximum at temperature of 48.7 $^{\circ}$C and a minimum at temperature of 47.6 $^{\circ}$C, reaching 3.68$\times$10$^{-9}$ A and -8.87$\times$10$^{-9}$, respectively. In the case of cubic sample, the recorded values of pyroelectric current are smaller (2.88$\times$10$^{-9}$ A when heated and -5.95$\times$10$^{-9}$ A when cooled). Temperatures of pyroelectric current extrema are shifted to 49.9 $ ^{\circ}$C (for heating) and 46.6 $^{\circ}$C (for cooling). The observed temperature hysteresis of the extreme values of pyroelectric current between heating and cooling process is 1.1 $^{\circ}$C for the plate-shaped sample and 3.3 $ ^{\circ}$C for the cubic sample. We expect that the longer path of heat transfer across the cubic sample in combination with different rates of heating (1.65 $^{\circ}$C/minute) and cooling 5.25 $^{\circ}$C/minute) are responsible for this difference. One may expect that in the case of cubic samples of larger dimensions this effect can be even more pronounced. 
\subsection{Pulse thermal excitation: $b$-sides are heated}
\label{sc:4}
\begin{figure}[ht]
	\centering
		\includegraphics[width=0.5\textwidth,angle=-90]{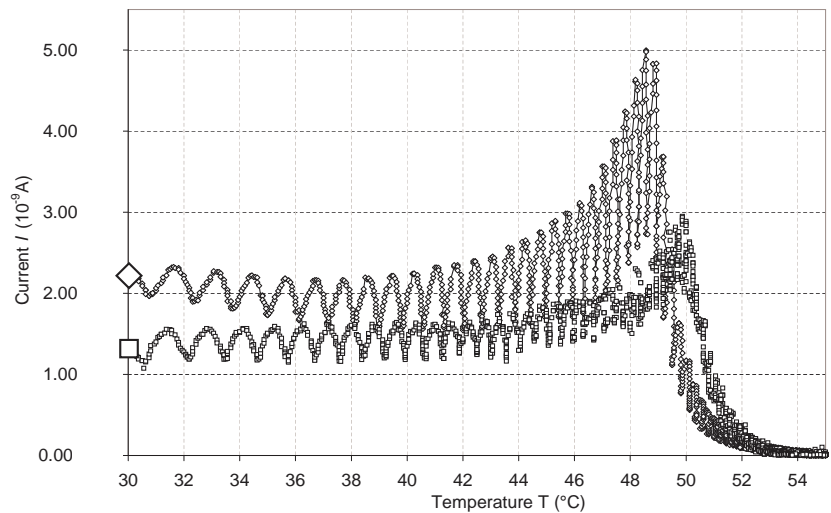}
	\caption{Pyroelectric current for plate-shaped ($\Diamond$) and cubic sample ($\Box$) for the pulse heating along $b$ axis.} 
	\label{fig:4}
\end{figure}

\begin{figure}[ht]
	\centering
		\includegraphics[width=0.5\textwidth,angle=-90]{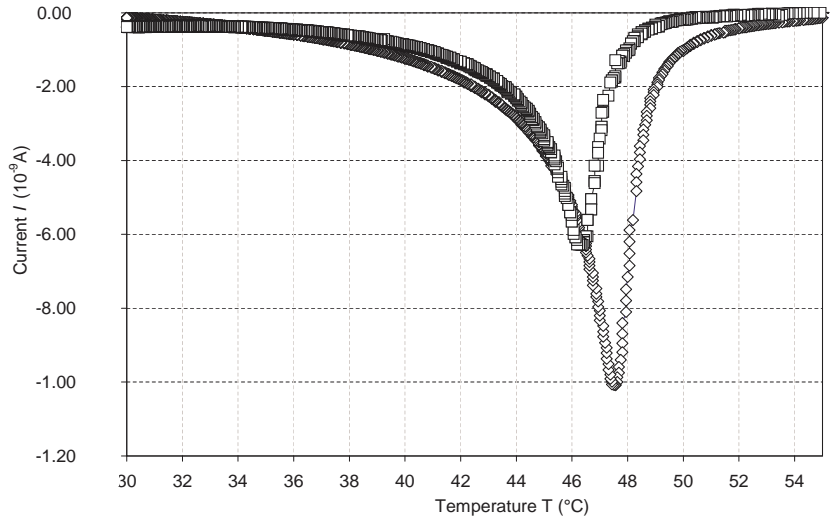}
	\caption{Pyroelectric current for plate-shaped ($\Diamond$) and cubic sample ($\Box$) for the cooling process after pulse heating applied to $b$-sides.} 
	\label{fig:5}
\end{figure}

To compare the dynamic response of cubic and plate-shaped samples, we conducted a test, namely the pulse measurements in the direction of $b$ axis. Thermal excitation in the form of waveform of 150 single rectangular  pulses of duration 5 s producing about 250 mJ energy in each pulse, was applied to the Peltier heaters situated in the sample holder. After each controlled pulse heating both plate-shaped and cubic samples were left in sample holder for spontaneous cooling down. Results of the experiment are presented in Figs. \ref{fig:4} and \ref{fig:5}.

Pyroelectric current of the plate-shaped sample in the pulse experiment has the global maximum at temperature 48.6$^{\circ}$C and minimum at 47.7$^{\circ}$C reaching 4.99×$\times$10$^{-9}$A and -9.84×$\times$10$^{-9}$A, respectively. Like in the linear excitation experiments, the cubic samples present smaller values of the pyroelectric current, namely maximum 2.93×$\times$10$^{-9}$A when heated and minimum -6.28×$\times$10$^{-9}$A when cooled. Temperatures of current maxima in the case of cubic sample are shifted to 49.9$^{\circ}$C and 46.3$^{\circ}$C, respectively. One may observe that values of pyroelectric response at the transition point are higher during pulse excitation, which we attribute to much greater temperature gradients created in the sample in comparison with the case of linear heating. Linear and spontaneous cooling down produce almost the same pyroelectric response in both the cubic and plate-shaped sample.

\subsection{Linear thermal excitation: $x, y$-sides are heated}
\label{sc:5}
Pyroelectric response to the linear excitation of the same parameters as described in section \ref{sc:3} was measured in the case of cubic samples in two orthogonal directions lying in {\it{b}}-plane. We denote such arbitrary selected measurements directions by the letters $x$ and $y$. It should further be noted that in our experiments the crystalline $a$ and $c$ axes were not precisely oriented.

\begin{figure}[ht]
	\centering
		\includegraphics[width=0.5\textwidth,angle=-90]{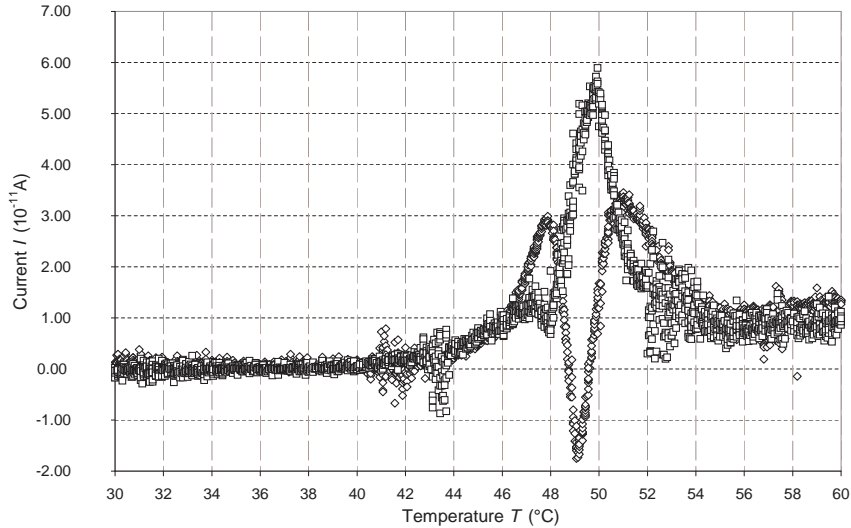}
	\caption{Pyroelectric current of cubic sample in $x$ ($\Diamond$) and $y$ ($\Box$) direction for the linear heating.} 
	\label{fig:6}
\end{figure}

\begin{figure}[ht]
	\centering
		\includegraphics[width=0.5\textwidth,angle=-90]{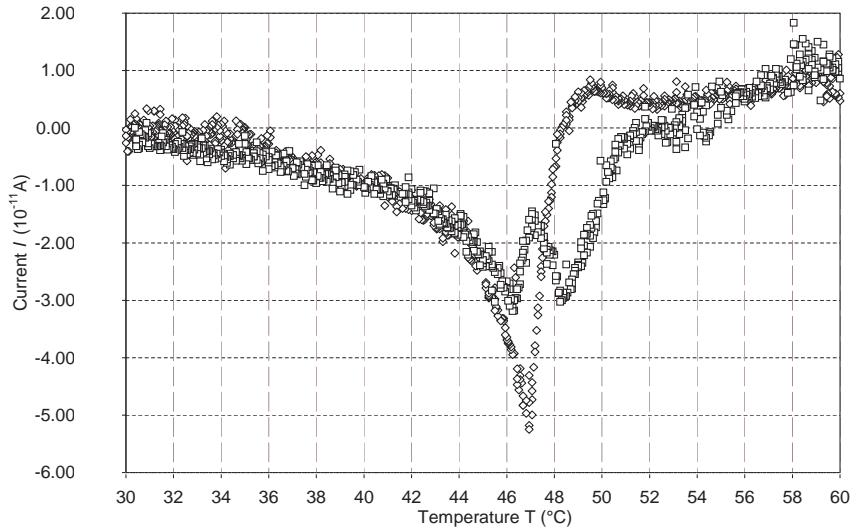}
	\caption{Pyroelectric current of cubic sample in $x$ ($\Diamond$) and $y$ ($\Box$) direction for the linear cooling.} 
	\label{fig:7}
\end{figure}

The results of measurements obtained for $x$ and $y$ directions for a cubic sample heated and cooled linearly with respective rates 1.65 $^{\circ}$C/minute (heating) and 5.25 $^{\circ}$C/minute (cooling) are presented in Figs. \ref{fig:6} and \ref{fig:7}.

Temperature characteristics of pyroelectric current in the case of cubic samples depend on the direction of observation ($x$ or $y$). During linear heating, pyroelectric current $I_x$ measured in $x$ direction has two maximal values of 2.95$\times$10$^{-11}$ A and 3.45$\times$10$^{-11}$ A at temperatures 47.8 $^{\circ}$C and 50.8 $^{\circ}$C, respectively. Pyroelectric current $I_y$ measured in $y$ direction has one maximum of value 5.89$\times$10$^{-11}$ A at temperature 49.9 $^{\circ}$C. During linear cooling down, one observes one minimum of $I_x$ of value -5.18$\times$10$^{-11}$ A at 47.0 $^{\circ}$C and two minima of $I_y$ of value -3.03$\times$10$^{-11}$ A and -3.19$\times$10$^{-11}$ A at respective temperatures 48.2 $^{\circ}$C and 46.3 $^{\circ}$C. Values of pyroelectric current measured in directions perpendicular to $b$ axis are more than 2.5 orders of magnitude smaller than those observed in $b$ direction. Extreme values of pyroelectric currents together with its temperatures measured in directions $b$, $x$ and $y$ are collected in Table \ref{table:1}. 

\begin{table}
\begin{adjustbox}{max width=0.6\textwidth}
\begin{tabular}{c|c|c|c|c|c|c}
\hline\hline
Measurement & Sample    &Excitation& Process & Curr.                  & Extrem.& Temp. of   \\
direction   &           &method    & type    & extrem.[A]             & type   & extrem. [{$^{\circ}$}C]\\
\hline
{\small}
$b$-axis    &plate      &pulse     &heating  & 4.99$\times 10^{-9}$   & max.   &48.6\\
            &shaped     &pulse     &cooling  &-9.84$\times 10^{-9}$   & min.   &47.5 \\
						\cline{3-7}
            &           &linear    &heating  & 3.98$\times 10^{-9}$   & max.   &48.7\\
            &           &linear    &cooling  &-8.87$\times 10^{-9}$   & min.   &47.6\\
\hline
$b$-axis    &cube       &pulse     &heating  & 2.93$\times 10^{-9}$   & max.   &49.9\\
            &           &pulse     &cooling  &-6.28$\times 10^{-9}$   & min.   &46.3\\
						\cline{3-7}
            &           &linear    &heating  & 2.88$\times 10^{-9}$   & max.   &49.9\\
            &           &linear    &cooling  &-5.95$\times 10^{-9}$   & min.   &46.6\\
\hline
$x$-axis    &cube       &linear    &heating  & 2.95$\times 10^{-11}$  & max.   &47.8\\
            &           &          &heating  &-1.75$\times 10^{-11}$  & min.   &49.1\\
            &           &          &heating  & 3.45$\times 10^{-11}$  & max.   &50.8\\
					  \cline{4-7}
            &           &          &cooling  &-5.18$\times 10^{-11}$  & min.   &46.9\\
\hline
$y$-axis    &cube       &linear    &heating  & 5.89$\times 10^{-11}$  & max.   &49.9\\
            \cline{4-7}
            &           &          &cooling  &-3.03$\times 10^{-11}$  & min.   &48.2\\
            &           &          &cooling  &-1.45$\times 10^{-11}$  & max.   &47.1\\
            &           &          &cooling  &-3.19$\times 10^{-11}$  & min.   &46.3\\
\hline \hline

\end{tabular}
\end{adjustbox}

\caption{Sequences of occurrence of extreme values of pyroelectric current for cubic and plate-shaped samples for heating and cooling.}
\label{table:1}
\end{table}

\section{Results and discussion}
\label{sc:6}
We investigated the pyroelectric response of cubic and plate-shaped samples of {\rm TGS} to pulse temperature excitations along ferroelectric $b$ axis. In both cases, we observed pyroelectric response to pulse excitation at temperatures far above the critical. This is the confirmation of results of Chynoweth \cite{Chynoweth} and our previous experiments \cite{Trybus}. To the best our knowledge, the mechanism of this phenomenon is not yet fully explained.

Pyroelectric current induced in {\rm TGS} single crystal in direction of ferroelectric $b$ axis is very sharp and rather strong in close proximity of $T_c$. This may obscure subtle phenomena which can be unmasked for other directions. This motivated us to perform similar experiments for cubic samples in two directions perpendicular to ferroelectric $b$ axis. We observed very weak signals -- more than 2.5 orders of magnitude weaker than those measured in $b$ axis direction at the same temperature conditions (linear heating and cooling). We discovered two correlated processes that seem to supplement each other.

Inspecting the data collected in Table \ref{table:1}, we note the following sequences of events for cubic samples:

\begin{enumerate}
\small
  \item When sample goes from ferroelectric to paraelectric phase (heating), for direction 
			\subitem $x$: local maximum (at 47.8 $^{\circ}$C) $\rightarrow$ local minimum (49.1 $^{\circ}$C) $\rightarrow$ local maximum (50.8 $^{\circ}$C), 
			\subitem $y$: global maximum (at 49.9 $^{\circ}$C),
			\subitem $b$: global maximum (at 49.9 $^{\circ}$C).

	\item When sample goes from paraelectric to ferroelectric phase (cooling), for direction:
			\subitem $x$: global minimum (at 46.9 $^{\circ}$C),
			\subitem $y$: local minimum (at 48.2 $^{\circ}$C) $\rightarrow$ local maximum (47.1 $^{\circ}$C)$\rightarrow$ local minimum (46.3 $^{\circ}$C), 
			\subitem $b$: global minimum (at 46.6 $^{\circ}$C).
\end{enumerate}

It can be concluded that the occurrence of global extreme values of pyroelectric currents, related to the phase transition and recorded along $b$ axis, are almost always preceded by local weak anomalies measured in directions $x$ and $y$ perpendicular to this axis.

\begin{figure}[ht]
	\centering
		\includegraphics[width=0.5\textwidth,angle=-90]{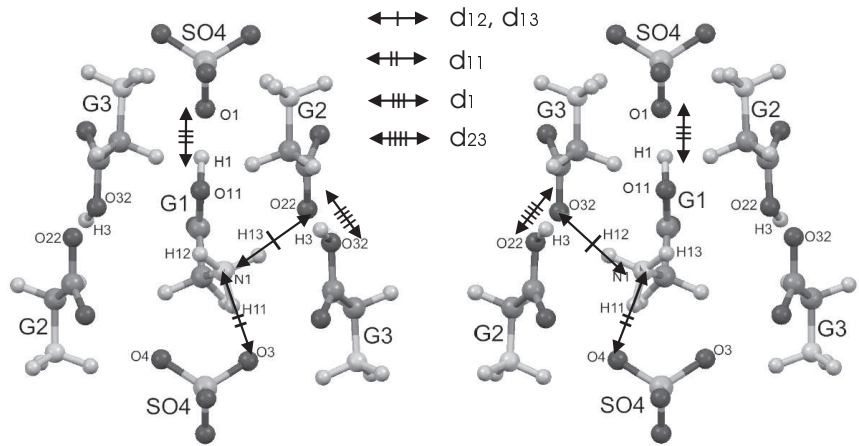}
	\caption{Possible configuration of hydrogen bonds between {\rm G1} and surrounding molecules and the short hydrogen bond {\rm G2}--{\rm G3} in ferroelectric phase \cite{Hudspeth, Goossens}. a) {\rm G1} in the left position, b) {\rm G1} in the right position.} 
	\label{fig:8}
\end{figure}

Hydrogen bonds between glycine molecules labeled {\rm G1},{\rm G2} and {\rm G3} as well as between {\rm G1} and {\rm SO}$_4$ molecules are depicted in Fig. \ref{fig:8}. Two glycine groups {\rm G2} and {\rm G3} are arranged almost perpendicular to the ferroelectric axis $b$. The length $d_{12}$ of the hydrogen bond {\rm G1}--{\rm G2} and the length $d_{13}$ of the bond {\rm G1}--{\rm G3} as well as the length $d_{23}$ of the bond {\rm G2}--{\rm G3} depend on temperature \cite{Hudspeth, Goossens}. The lengths $d_{12}$ and $d_{13}$ increase, whilst $d_{23}$ decreases with growing temperature. Lengths $d_{11}$ and $d_1$ of hydrogen bonds {\rm N1}--{\rm O3} and {\rm N1}--{\rm O4} as well as {\rm O11}--{\rm O1} increase with the rise of temperature.

As the hydrogen bond between {\rm G1} and {\rm SO}$_4$ gives very weak contribution to the polarization along $b$ axis, we do not mention it in the following discussion.

In the paraelectric phase ($T>T_c$), lengths $d_{12}$ and $d_{13}$ are sufficiently large, so molecules {\rm G2} and {\rm G3} do not influence the {\rm G1} molecules. The energy of thermal motion makes the probability of choosing each of two symmetrical positions of nitrogen atom of {\rm G1} group the same. Hence, the orientation of {\rm NH}$_3$ group is disordered across the mirror plane. Therefore, the {\rm TGS} crystal has the space group symmetry $P2_{1}/m$. No spontaneous polarization is present. 

When temperature diminishes below $T_c$, the length $d_{23}$ increases. As a result, {\rm N1} atom may develop two equivalent bonds with oxygens {\rm O22} or {\rm O32}. In the diffuse scattering of X-rays experiments, it was found that the position of {\rm H}3 atom in {\rm G2}--{\rm G3} bond is correlated with the orientation of {\rm G1} molecule \cite{Hudspeth,Goossens}. In the case of the right position of {\rm NH}$_3$ group, {\rm H3} atom is covalently bonded with {\rm O32} atom, and {\rm N1} is hydrogen bonded to {\rm O22}. In the case of the left position of {\rm NH}$_3$, the {\rm H3} is bonded with atom {\rm O22}, while {\rm N1} creates hydrogen bond with {\rm O32}. 

Hudspeth et al. \cite{Goossens} underlined that the dominant mechanism of interaction of {\rm G1} molecules is indirect via hydrogen bonds {\rm G1}--{\rm G2} or {\rm G1}--{\rm G3}, giving the zigzag networks of interactions as illustrated in Fig. 6 of the paper by Hudspeth et al. \cite{Goossens}. As a result, the mirror plane disappears, the {\rm TGS} structure has space group symmetry $P2_1$, and the crystal becomes polarized along the $b$ axis. 

The reverse process can be observed when crystal approaches $T_c$ transitioning from ferroelectric to paraelectric phase. The length $d_{23}$ decreases, whilst lengths $d_{12}$ and $d_{13}$ increase. The well of potential with {\rm NH}$_3$ particle inside becomes lower. Increasing thermal vibrations make right and left position of {\rm NH}$_3$ equivalent. The long range order between {\rm G1} molecules disappears. The {\rm NH}$_3$ group becomes disordered across the mirror plane, and the crystal space group symmetry becomes $P2_{1}/m$. Spontaneous polarization disappears.

Each of the above mentioned hydrogen bonds adds components of polarization vector along the $b$ and $x$ and $y$ axes. These components are very weak in comparison with the component of polarization vector induced by nitrogen {\rm{N1}} atoms. Therefore, the only way to register such weak signals is to measure them in plane perpendicular to the polar axis. Non-vanishing currents $I_x$ and $I_y$ measured in the mutually perpendicular directions perpendicular to polar $b$ axis show that the listed hydrogen bonds are activated at different temperatures and these components of the induced polarization vectors have various orientations.

We believe that our results are in agreement with the observations of Hudspeth \cite{Hudspeth} and Hudspeth et al. \cite{Goossens}. Our method seems to be a good tool for the disclosure of weak effects accompanying the phase transition in {\rm TGS}.


\newpage
\begin{thebibliography}{30}
\bibitem{Wood} E.A. Wood, A.N. Holden, Acta Crystallogr. 10, 145 (1957).
\bibitem{Trybus} M. Trybus, W. Proszak, B. Wo{\'s}, Infrared Phys.Techn. 61, 81 (2013). 
\bibitem{Wos} M. Trybus, B. Wo{\'s}, Infrared Phys. Techn. 71, 526 (2015).
\bibitem{Chynoweth} A.G. Chynoweth, Phys. Rev. 111, 1235 (1960).
\bibitem{White} D.J. White, H.H. Wieder, J. Appl. Phys. 34, 2847 (1963).
\bibitem{Shaulov} A. Shaulov, M. Simhony, J. Appl. Phys. 43, 1440 (1972).
\bibitem{Simhony} M. Simhony, A. Shaulov and A. Maman, J. Appl. Phys. 44, 2464 (1973). 
\bibitem{Hadni} A. Hadni, R. Thomas, Ferroelectrics 4, 39 (1972). 
\bibitem{Lambert} A. Hadni, J.P. Lambert, M.M. Pradhan and R. Thomas, Infrared Physics 13, 305 (1973). 
\bibitem{Bhide} V.G. Bhide, M.M. Pradhan and R.K. Garg, Pramana 8, 276 (1977). 
\bibitem{Al-Allak} H.M. Al-Allak, R. Dewsberry, J. Mater. Sciences 27, 6743 (1992). 
\bibitem{Fugiel-a} B. Fugiel, Solid State Comm. 1222, 2372 (2002). 
\bibitem{Fugiel-b} B. Fugiel, Ferroelectrics 330, 45 (2006).
\bibitem{Cwikiel-a} B. Fugiel, K. {\'Cw}ikiel and W. Serweci{\'n}ski, Journ. Phys: Condens. Matter 141, 11837 (2000).
\bibitem{Correira} A. Correira, J. Massanell, N. Garica, A.P. Levanyuk, A. Zlatkin and J. Przes{\l{}}awski, Appl. Phys. Lett. 68, 2796 (1996). 
\bibitem{Bluhm} H. Bluhm, R. Wissendanger and K-P. Meyer, Surface. J. Vac. Sci. Technol. B14, 1180 (1996).  
\bibitem{Abplanalp} M. Abplanalp, L.M. Eng and P. G{\"u}nter, Appl. Phys. A 66, S231 (1998). 
\bibitem{Orlik} X.K. Orlik, V. Likodimos, L. Pardi, M. Labardi and M. Allegrini, App. Phys. Lett. 70, 1321 (2000). 
\bibitem{Shin} S. Shin, J. Baek, J.W. Hong and Z.G. Khim, J. Appl. Phys. 96, 4372 (2004).
\bibitem{Cwikiel-b} K. {\'C}wikiel, D. Kajewski, Phase Transit. 83, 595 (2010). 
\bibitem{Hudspeth} J. M. Hudspeth, Ph.D. thesis, Australian National University, 2012.
\bibitem{Goossens} J.M. Hudspeth, D.J. Goossens, T.R. Wellbery and M.J. Gutmann, J. Matter. Sci. 48, 6605 (2013).

\end{thebibliography}
\end{document}